\documentclass[12pt,a4paper]{article}

\usepackage{ifthen} % for conditional statements
\newboolean{pdflatex}
\setboolean{pdflatex}{true} % False for eps figures 
\newboolean{articletitles}
\setboolean{articletitles}{true} % False removes titles in references
\newboolean{uprightparticles}
\setboolean{uprightparticles}{false} %True for upright particle 
\newboolean{inbibliography}
\setboolean{inbibliography}{false} %True once you enter the bibliography
\usepackage{multirow}
\usepackage{xspace} % To avoid problems with missing or double spaces after

\def\paperauthors{LHCb collaboration} % Leave as is for PAPER and CONF
\def\paperasciititle{Charmonium production in  
SQRT(sNN)=68.5 GeV pNe collisions} % Set ASCII title here
\def\papertitle{Charmonium  production  in $p$Ne collisions at $\sqsnn=68.5\gev$  }
\def\paperkeywords{{High Energy Physics}, {LHCb}} % Comma separated list
\def\papercopyright{2021 CERN for the benefit of the LHCb collaboration}
\def\paperlicence{CC-BY-4.0}
\def\paperlicenceurl{https://creativecommons.org/licenses/by/4.0/}

\usepackage[top=1in, bottom=1.25in, left=1in, right=1in]{geometry}
\usepackage{microtype}
\usepackage{lineno}  % for line numbering during review
\usepackage{caption} %these three command get the figure and table captions automatically small

\usepackage{graphicx}  % to include figures (can also use other packages)
\usepackage{color}
\usepackage{colortbl}
\graphicspath{{./figs/}} % Make Latex search fig subdir for figures

\usepackage{amsmath,amssymb,amsfonts} % Adds a large collection of math symbols
\usepackage{upgreek} % Adds in support for greek letters in roman typeset

\newcommand*\patchAmsMathEnvironmentForLineno[1]{%
\expandafter\let\csname old#1\expandafter\endcsname\csname #1\endcsname
\expandafter\let\csname oldend#1\expandafter\endcsname\csname
end#1\endcsname
 \renewenvironment{#1}%
   {\linenomath\csname old#1\endcsname}%
   {\csname oldend#1\endcsname\endlinenomath}%
}
\newcommand*\patchBothAmsMathEnvironmentsForLineno[1]{%
  \patchAmsMathEnvironmentForLineno{#1}%
  \patchAmsMathEnvironmentForLineno{#1*}%
}
\AtBeginDocument{%
\patchBothAmsMathEnvironmentsForLineno{equation}%
\patchBothAmsMathEnvironmentsForLineno{align}%
\patchBothAmsMathEnvironmentsForLineno{flalign}%
\patchBothAmsMathEnvironmentsForLineno{alignat}%
\patchBothAmsMathEnvironmentsForLineno{gather}%
\patchBothAmsMathEnvironmentsForLineno{multline}%
\patchBothAmsMathEnvironmentsForLineno{eqnarray}%
}

\usepackage{hyperxmp}

\usepackage[pdftex,
            pdfauthor={\paperauthors},
            pdftitle={\paperasciititle},
            pdfkeywords={\paperkeywords},
            pdfcopyright={Copyright (C) \papercopyright},
            pdflicenseurl={\paperlicenceurl}]{hyperref}

\usepackage[all]{hypcap} % Internal hyperlinks to floats.

\usepackage{upgreek}

\usepackage{xspace} 
\usepackage{upgreek}

\def\lhcb   {\mbox{LHCb}\xspace}

\def\MagUp {\mbox{\em Mag\kern -0.05em Up}\xspace}

\ifthenelse{\boolean{uprightparticles}}%
{

 \def\Pmu         {\ensuremath{\upmu}\xspace}

 \def\Ppi         {\ensuremath{\uppi}\xspace}

 \def\Ppsi        {\ensuremath{\uppsi}\xspace}

 \def\PDelta      {\ensuremath{\Delta}\xspace}                 
 \def\PXi         {\ensuremath{\Xi}\xspace}                 
 \def\PLambda     {\ensuremath{\Lambda}\xspace}                 
 \def\PSigma      {\ensuremath{\Sigma}\xspace}                 
 \def\POmega      {\ensuremath{\Omega}\xspace}                 
 \def\PUpsilon    {\ensuremath{\Upsilon}\xspace}
 \let\oldPi\Pi
 \def\PPi         {\ensuremath{\oldPi}\xspace}

 \def\PB      {\ensuremath{\mathrm{B}}\xspace}                 
                  
 \def\PD      {\ensuremath{\mathrm{D}}\xspace}

 \def\PJ      {\ensuremath{\mathrm{J}}\xspace}                 
 \def\PK      {\ensuremath{\mathrm{K}}\xspace}

 \def\Pb      {\ensuremath{\mathrm{b}}\xspace}                 
 \def\Pc      {\ensuremath{\mathrm{c}}\xspace}

 \def\Pi      {\ensuremath{\mathrm{i}}\xspace}

 \def\Ps      {\ensuremath{\mathrm{s}}\xspace}

 \def\thebaroffset{0.0em}
}
{

 \def\Pmu         {\ensuremath{\mu}\xspace}

 \def\Ppi         {\ensuremath{\pi}\xspace}

 \def\Ppsi        {\ensuremath{\psi}\xspace}                 
                  
 \mathchardef\PDelta="7101
 \mathchardef\PXi="7104
 \mathchardef\PLambda="7103
 \mathchardef\PSigma="7106
 \mathchardef\POmega="710A
 \mathchardef\PUpsilon="7107
 \mathchardef\PPi="7105
                  
 \def\PB      {\ensuremath{B}\xspace}                 
                  
 \def\PD      {\ensuremath{D}\xspace}

 \def\PJ      {\ensuremath{J}\xspace}                 
 \def\PK      {\ensuremath{K}\xspace}

 \def\Pb      {\ensuremath{b}\xspace}                 
 \def\Pc      {\ensuremath{c}\xspace}

 \def\Pi      {\ensuremath{i}\xspace}

 \def\Ps      {\ensuremath{s}\xspace}

 \def\thebaroffset{0.18em}
}
\newcommand{\offsetoverline}[2][\thebaroffset]{\kern #1\overline{\kern -#1 #2}}%

\makeatletter
\ifcase \@ptsize \relax% 10pt
  \newcommand{\miniscule}{\@setfontsize\miniscule{4}{5}}% \tiny: 5/6
\or% 11pt
  \newcommand{\miniscule}{\@setfontsize\miniscule{5}{6}}% \tiny: 6/7
\or% 12pt
  \newcommand{\miniscule}{\@setfontsize\miniscule{5}{6}}% \tiny: 6/7
\fi
\makeatother

\DeclareRobustCommand{\optbar}[1]{\shortstack{{\miniscule (\rule[.5ex]{1.25em}{.18mm})}
  \\ [-.7ex] $#1$}}

\def\mup        {{\ensuremath{\Pmu^+}}\xspace}
\def\mun        {{\ensuremath{\Pmu^-}}\xspace} % muon negative (\mum is taken)

\def\squark    {{\ensuremath{\Ps}}\xspace}

\def\cquark    {{\ensuremath{\Pc}}\xspace}
\def\cquarkbar {{\ensuremath{\overline \cquark}}\xspace}
\def\ccbar     {{\ensuremath{\cquark\cquarkbar}}\xspace}
\def\bquark    {{\ensuremath{\Pb}}\xspace}

\def\pion   {{\ensuremath{\Ppi}}\xspace}

\def\pip    {{\ensuremath{\pion^+}}\xspace}

\def\kaon    {{\ensuremath{\PK}}\xspace}
\def\KorKbar {\kern \thebaroffset\optbar{\kern -\thebaroffset \PK}{}\xspace}

\def\Km      {{\ensuremath{\kaon^-}}\xspace}

\def\D       {{\ensuremath{\PD}}\xspace}

\def\DorDbar {\kern \thebaroffset\optbar{\kern -\thebaroffset \PD}\xspace}
\def\Dz      {{\ensuremath{\D^0}}\xspace}

\def\Dp      {{\ensuremath{\D^+}}\xspace}
\def\Dm      {{\ensuremath{\D^-}}\xspace}

\def\DpDm    {\ensuremath{\Dp {\kern -0.16em \Dm}}\xspace}

\def\B       {{\ensuremath{\PB}}\xspace}

\def\BorBbar {\kern \thebaroffset\optbar{\kern -\thebaroffset \PB}\xspace}

\def\Bd      {{\ensuremath{\B^0}}\xspace}

\def\BdorBdbar {\kern \thebaroffset\optbar{\kern -\thebaroffset \Bd}\xspace}

\def\Bs      {{\ensuremath{\B^0_\squark}}\xspace}

\def\BsorBsbar {\kern \thebaroffset\optbar{\kern -\thebaroffset \Bs}\xspace}

\def\jpsi     {{\ensuremath{{\PJ\mskip -3mu/\mskip -2mu\Ppsi}}}\xspace}
\def\psitwos  {{\ensuremath{\Ppsi{(2S)}}}\xspace}

\def\Upsilonres  {{\ensuremath{\PUpsilon}}\xspace}
\def\Y#1S{\ensuremath{\PUpsilon{(#1S)}}\xspace}

\def\LorLbar     {\kern \thebaroffset\optbar{\kern -\thebaroffset \PLambda}\xspace}

\def\BF         {{\ensuremath{\mathcal{B}}}\xspace}

\def\to                 {\ensuremath{\rightarrow}\xspace}

\def\AT#1     {\ensuremath{A_{\mathrm{T}}^{#1}}\xspace}           % 2

\def\C#1      {\ensuremath{\mathcal{C}_{#1}}\xspace}                       % 9
\def\Cp#1     {\ensuremath{\mathcal{C}_{#1}^{'}}\xspace}                    % 7
\def\Ceff#1   {\ensuremath{\mathcal{C}_{#1}^{\mathrm{(eff)}}}\xspace}        % 9  
\def\Cpeff#1  {\ensuremath{\mathcal{C}_{#1}^{'\mathrm{(eff)}}}\xspace}       % 7
\def\Ope#1    {\ensuremath{\mathcal{O}_{#1}}\xspace}                       % 2
\def\Opep#1   {\ensuremath{\mathcal{O}_{#1}^{'}}\xspace}                    % 7

\newcommand{\aunit}[1]{\ensuremath{\text{\,#1}}}       
\newcommand{\tev}{\aunit{Te\kern -0.1em V}\xspace}
\newcommand{\gev}{\aunit{Ge\kern -0.1em V}\xspace}
\newcommand{\mev}{\aunit{Me\kern -0.1em V}\xspace}
\newcommand{\kev}{\aunit{ke\kern -0.1em V}\xspace}
\newcommand{\ev}{\aunit{e\kern -0.1em V}\xspace}
 
\newcommand{\mevc}{\ensuremath{\aunit{Me\kern -0.1em V\!/}c}\xspace}
\newcommand{\gevc}{\ensuremath{\aunit{Ge\kern -0.1em V\!/}c}\xspace}
\newcommand{\mevcc}{\ensuremath{\aunit{Me\kern -0.1em V\!/}c^2}\xspace}
\newcommand{\gevcc}{\ensuremath{\aunit{Ge\kern -0.1em V\!/}c^2}\xspace}
\def\gsim{{~\raise.15em\hbox{$>$}\kern-.85em
          \lower.35em\hbox{$\sim$}~}\xspace}
\def\lsim{{~\raise.15em\hbox{$<$}\kern-.85em
          \lower.35em\hbox{$\sim$}~}\xspace}

\def\sqsnn {\ensuremath{\protect\sqrt{s_{\scriptscriptstyle\text{NN}}}}\xspace}
\def\pt         {\ensuremath{p_{\mathrm{T}}}\xspace}

\def\evtgen     {\mbox{\textsc{EvtGen}}\xspace}

\def\geant      {\mbox{\textsc{Geant4}}\xspace}

\def\photos     {\mbox{\textsc{Photos}}\xspace}

\def\pythia     {\mbox{\textsc{Pythia}}\xspace}

\def\tell1  {TELL1\xspace}
\def\ukl1   {UKL1\xspace}

\newcommand{\lhcborcid}[1]{\href{https://orcid.org/#1}{\hspace*{0.1em}\raisebox{-0.45ex}{\includegraphics[width=1em]{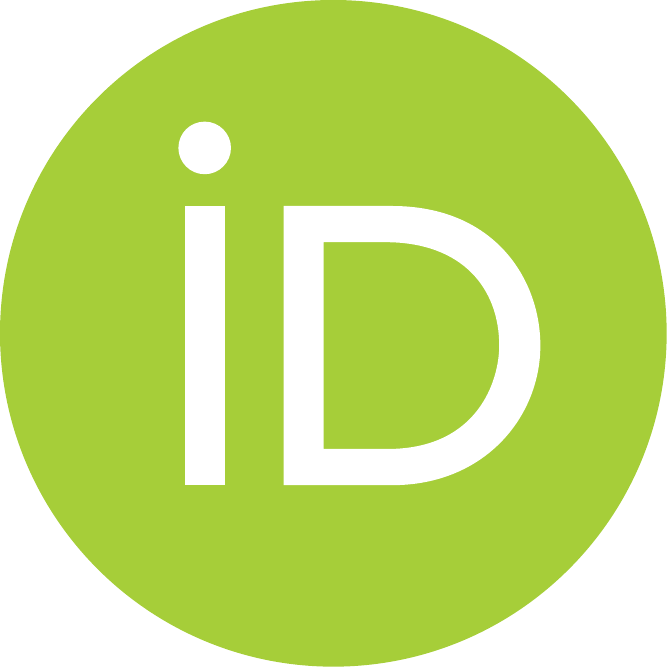}}}}

\def\epos      {\mbox{\textsc{Epos}}\xspace}

\newcommand{\pp}{\ensuremath{pp}\xspace}
\newcommand{\pHe}{\ensuremath{p\text{He}}\xspace}

\newcommand{\pNe}{\ensuremath{p\text{Ne}}\xspace}

\let\oldequation\equation
\let\oldendequation\endequation

\renewenvironment{equation}
  {\linenomathNonumbers\oldequation}
  {\oldendequation\endlinenomath}

\usepackage{cite} % Allows for ranges in citations
\usepackage{mciteplus}

\begin{document}

%%%%%%%%%%%%%%%%%%%%%%%%%
%%%%% Title     %%%%%%%%%
%%%%%%%%%%%%%%%%%%%%%%%%%
\renewcommand{\thefootnote}{\fnsymbol{footnote}}
\setcounter{footnote}{1}

%%%%%%%%%%%%%%%%%%%%%%%%%
%%%%%  TITLE PAGE  %%%%%%
%%%%%%%%%%%%%%%%%%%%%%%%%
\begin{titlepage}
\pagenumbering{roman}
% Header ---------------------------------------------------
\vspace*{-1.5cm}
\centerline{\large EUROPEAN ORGANIZATION FOR NUCLEAR RESEARCH (CERN)}
\vspace*{1.5cm}
\noindent
\begin{tabular*}{\linewidth}{lc@{\extracolsep{\fill}}r@{\extracolsep{0pt}}}
\ifthenelse{\boolean{pdflatex}}% Logo format choice
{\vspace*{-1.5cm}\mbox{\!\!\!\includegraphics[width=.14\textwidth]{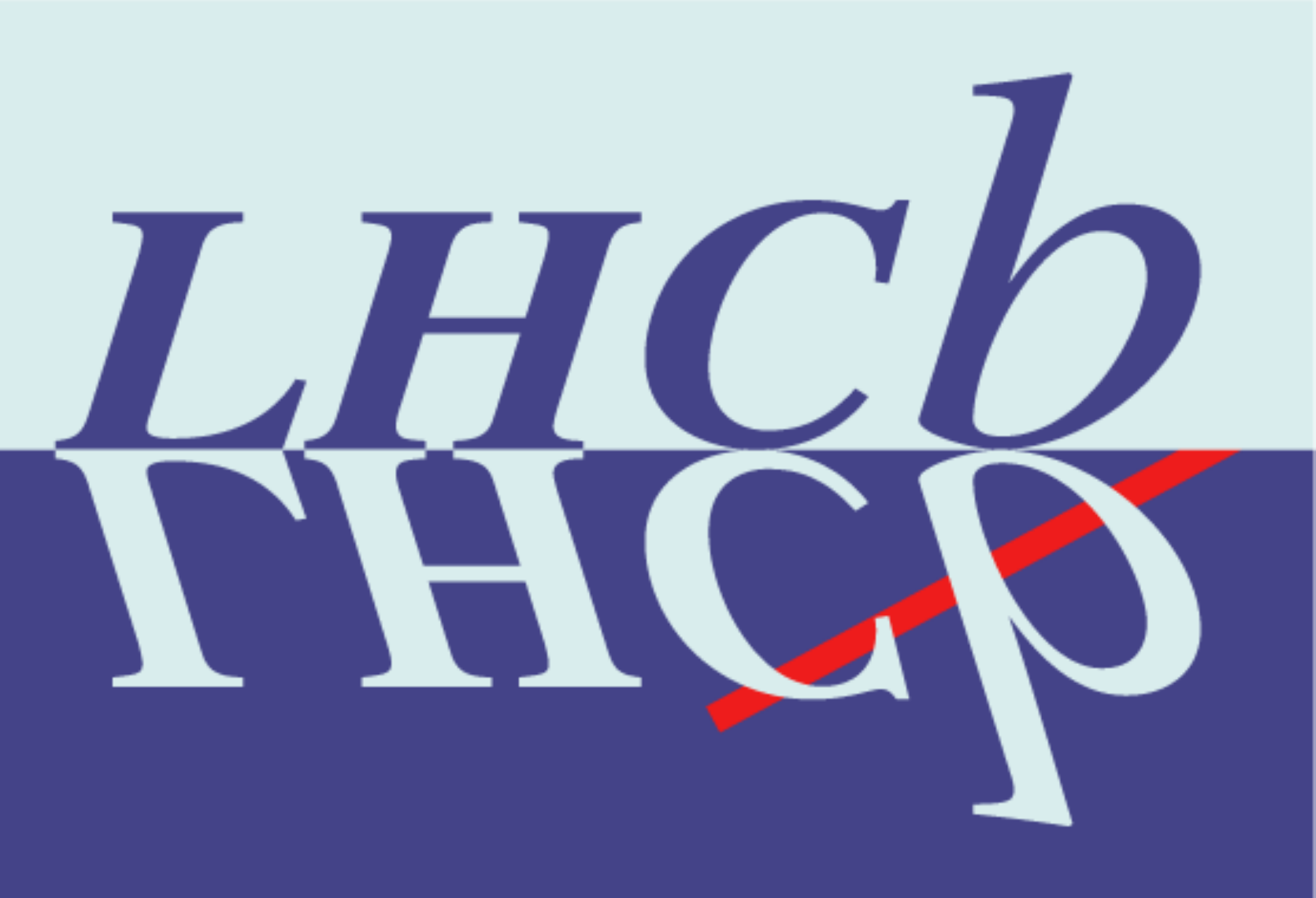}} & &}%
{\vspace*{-1.2cm}\mbox{\!\!\!\includegraphics[width=.12\textwidth]{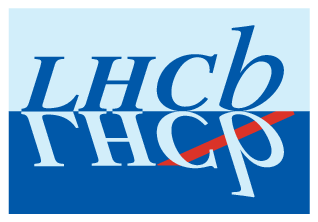}} & &}%
\\
 & & CERN-EP-2022-216 \\  % ID 
 & & LHCb-PAPER-2022-014 \\  % ID 
 & & August 10, 2023 \\ % Date - Can also hardwire e.g.: 23 March 2010
 & & \\
% not in paper \hline
\end{tabular*}
\vspace*{4.0cm}
% Title --------------------------------------------------
{\normalfont\bfseries\boldmath\huge
\begin{center}
% DO NOT EDIT HERE. Instead edit macro in main.tex to keep metadata correct
  \papertitle 
\end{center}
}
\vspace*{2.0cm}
% Authors -------------------------------------------------
\begin{center}
%In the footnote, replace 'paper' by 'Letter' in case of submission to PRL or PLB 
% Edit macro in main.tex to keep metadata correct
\paperauthors\footnote{Authors are listed at the end of this paper.}
\end{center}
\vspace{\fill}
% Abstract -----------------------------------------------
\begin{abstract}
  \noindent
The measurement of charmonium states produced in proton-neon (\pNe) collisions by the LHCb experiment in its fixed-target configuration is presented. The production of \jpsi and \psitwos mesons is studied with a beam of 2.5\tev protons colliding on gaseous neon targets at rest, corresponding to a nucleon-nucleon centre-of-mass energy $\sqsnn=68.5\gev$. The data sample corresponds to an integrated luminosity of $21.7\pm 1.4 $ nb$^{-1}$. The \jpsi and \psitwos hadrons are reconstructed in \mup\mun final states. The \jpsi production cross-section per target nucleon in the centre-of-mass rapidity range $y^{*}\in [-2.29, 0]$ is found to be $506 \pm 8 \pm  46 \textrm{ nb/nucleon}$. 
The ratio of \jpsi and \Dz cross-sections is evaluated to $(1.06 \pm 0.02 \pm 0.09)\%$.
The \psitwos to \jpsi relative production rate is found to be $(1.67 \pm 0.27\pm 0.10)\%$ in good agreement with other measurements involving beam and target nuclei of similar sizes.
\end{abstract}
\vspace*{2.0cm}
\begin{center}
Published in Eur. Phys. J. \textbf{C83} (2023) 625.
\end{center}
\vspace{\fill}
{\footnotesize 
% Edit macro in main.tex to keep metadata correct
\centerline{\copyright~\papercopyright. \href{\paperlicenceurl}{\paperlicence} licence.}}
\vspace*{2mm}
\end{titlepage}

%%%%%%%%%%%%%%%%%%%%%%%%%%%%%%%%
%%%%%  EOD OF TITLE PAGE  %%%%%%
%%%%%%%%%%%%%%%%%%%%%%%%%%%%%%%%

%  empty page follows the title page ----
\newpage
\setcounter{page}{2}
\mbox{~}
\cleardoublepage

\renewcommand{\thefootnote}{\arabic{footnote}}
\setcounter{footnote}{0}

%%%%%%%%%%%%%%%%%%%%%%%%%%%%%%%%
%%%%%  Table of Content   %%%%%%
%%%%%%%%%%%%%%%%%%%%%%%%%%%%%%%%
%%%% Uncomment next 2 lines if desired
%\tableofcontents
%\cleardoublepage

%%%%%%%%%%%%%%%%%%%%%%%%%
%%%%% Main text %%%%%%%%%
%%%%%%%%%%%%%%%%%%%%%%%%%

\pagestyle{plain} % restore page numbers for the main text
\setcounter{page}{1}
\pagenumbering{arabic}

%% Uncomment during review phase. 
%% Comment before a final submission.
%\linenumbers

%%%%% Introduction 
The production of charmonia, \ccbar bound states, is interesting to study  in proton-proton, proton-nucleus and nucleus-nucleus collisions. This process involves two scales: that of the \ccbar pair production, which can be studied in proton-proton collisions; and that of hadronization, for which proton-nucleus collisions can bring decisive insights.

Several initial- and final-state effects occur in proton-nucleus collisions that can modify charmonium production with respect to proton-proton collisions. Charmonium production can be suppressed by nuclear absorption \cite{NAPLB207} and can be affected by multiple scattering \cite{PRD77}, and energy loss by radiation \cite{ELJHEP03} in the proton-nucleus overlapping region. Charmonium states can also be dissociated by comovers \cite{CapellaPRL85} or  affected by the modification, namely shadowing or anti-shadowing, of the parton flux inside the nucleus \cite{PDFEPJC77,PDFPRD93}. These so-called cold nuclear-matter effects (CNM) depend on the collision energy, the transverse momentum and rapidity of the produced charmonium state, as well as the size of the target nucleus. It is therefore essential to carry out charmonium measurements over a wide range of experimental conditions. 
Moreover, the understanding of charmonium production and hadronization mechanisms can be significantly improved by comparison with measurements of the overall charm quark production, for which \Dz mesons are a good proxy, as their production dominates over other charm hadrons. 

In this paper, a measurement of charmonium production in the \lhcb fixed-target configuration is presented. The production of \jpsi mesons is studied in collisions of protons with energies of 2.5 TeV incident on neon  nuclei at rest, resulting in centre-of-mass energies of $\sqsnn=68.5\gev$. It is also compared with the production of $D^0$ mesons measured in the same conditions \cite{D0pNe}. In addition, the first measurement of the relative production rate of \psitwos and \jpsi mesons in this fixed-target configuration is reported. 

%%%% Detector
The LHCb detector~\cite{Alves:2008zz, LHCb-DP-2014-002} is a single-arm forward spectrometer covering the pseudorapidity range $2 < \eta < 5$. It was designed primarly for the study of particles containing $c$ or $b$ quarks.
The main detector elements are: the silicon-strip vertex locator (VELO) surrounding the interaction region that allows to precisely reconstruct the decay vertex of \cquark and \bquark hadrons; a tracking system with a warm magnet and tracking stations that provide a measurement of the momentum of charged particles; two ring-imaging Cherenkov detectors that provide  discrimination between different species of charged hadrons; a calorimeter system consisting of scintillating-pad and preshower detectors in front of the electromagnetic and hadronic calorimeters; and a muon detector composed of alternating layers of iron and multiwire proportional chambers. The system for measuring the overlap with gas (SMOG) \cite{LHCb-PAPER-2014-047,Barschel} 
is used to measure LHC beam profiles. It enables the injection of gases with pressure of $O(10^{-7})$ mbar in the beam-pipe section inside the VELO, allowing LHCb to operate as a fixed-target experiment. SMOG allows the injection of noble gases and therefore gives the unique opportunity to study nucleus-nucleus and proton-nucleus collisions on various targets. Due to the boost induced by the high-energy proton beam, the LHCb acceptance covers the backward
rapidity hemisphere in the nucleon-nucleon centre-of-mass system of the reaction, $-2.29<y^\star<0$.

%%%% Trigger
Events are selected by the two-stage trigger system \cite{trigger}. The  first level is implemented in hardware and uses information provided by the calorimeters
and the muon detectors, while the second is a software trigger. 
The hardware trigger requires at least one identified muon for the reconstruction of the $\jpsi \to \mup \mun $ and  $\psitwos \to \mup \mun$ decays. The software trigger requires two well-reconstructed muons having an invariant mass, $m_{\mup\mun}$, greater than $2700\mevcc$. 

%%%%% Data
The data samples correspond to a collider configuration in which proton bunches moving towards the detector do not cross any bunch moving in the opposite direction. Unlike in proton-proton (\pp) collisions, no nominal interaction point exists in the fixed-target case. Therefore, events are required to have a reconstructed primary vertex (PV) with its coordinate along the beam axis ($z$) being within the fiducial region \mbox{$z_{PV}\in[-200,-100]\cup[100,150]$~mm} (where $z_{PV} = 0$~mm is the nominal position of the \pp interaction point), within which high reconstruction efficiencies are achieved and calibration samples are available. Residual \pp collision events, are suppressed by vetoing events with activity in the backward direction with respect to the beam direction, based on the number of hits in VELO stations upstream of the interaction region. The region $-100 < z_{PV} < 100$ mm, where most of the residual \pp collisions occur, is also vetoed.
\begin{figure}[t]
\begin{center}
\includegraphics[width=0.49\textwidth]{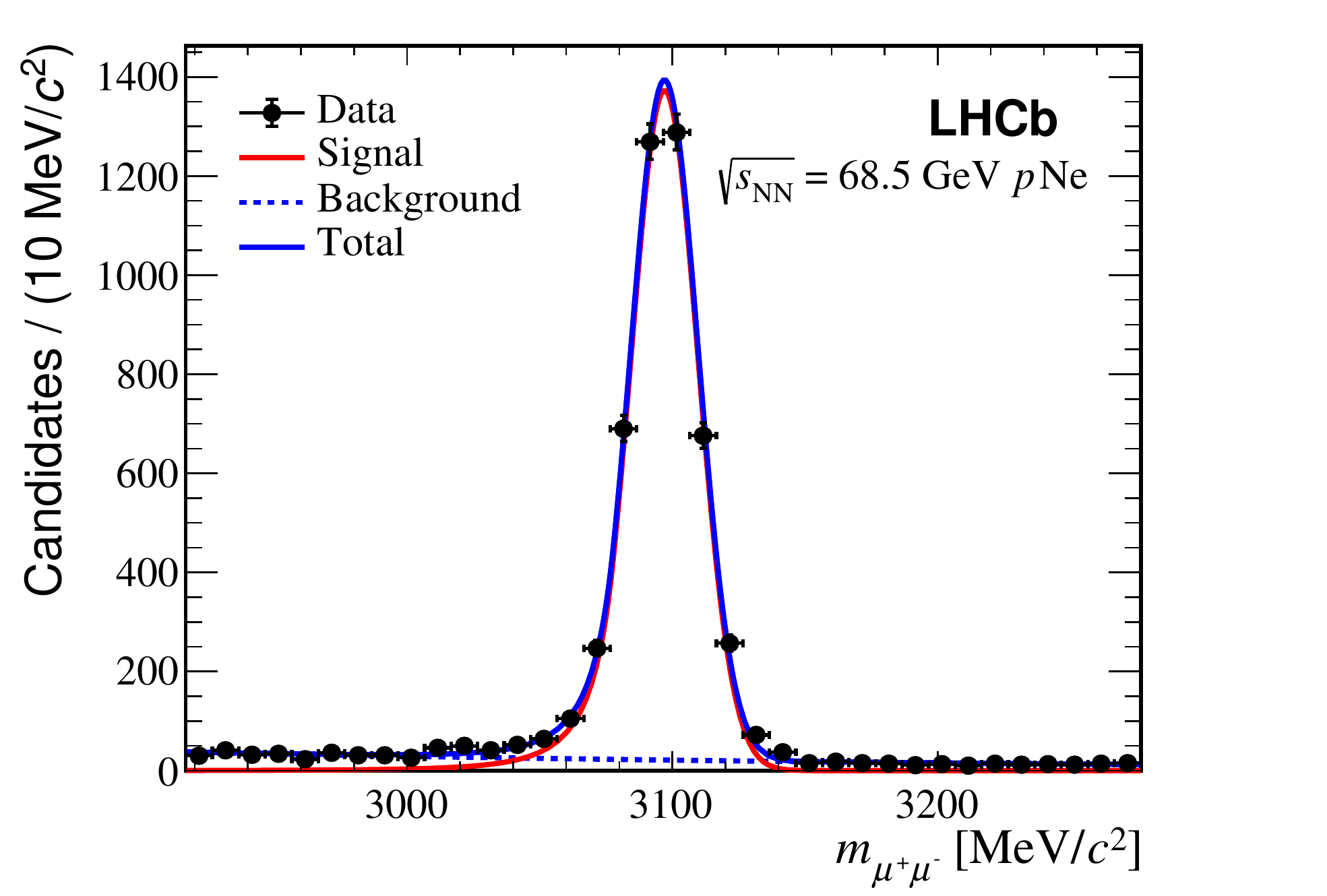}
\includegraphics[width=0.49\textwidth]{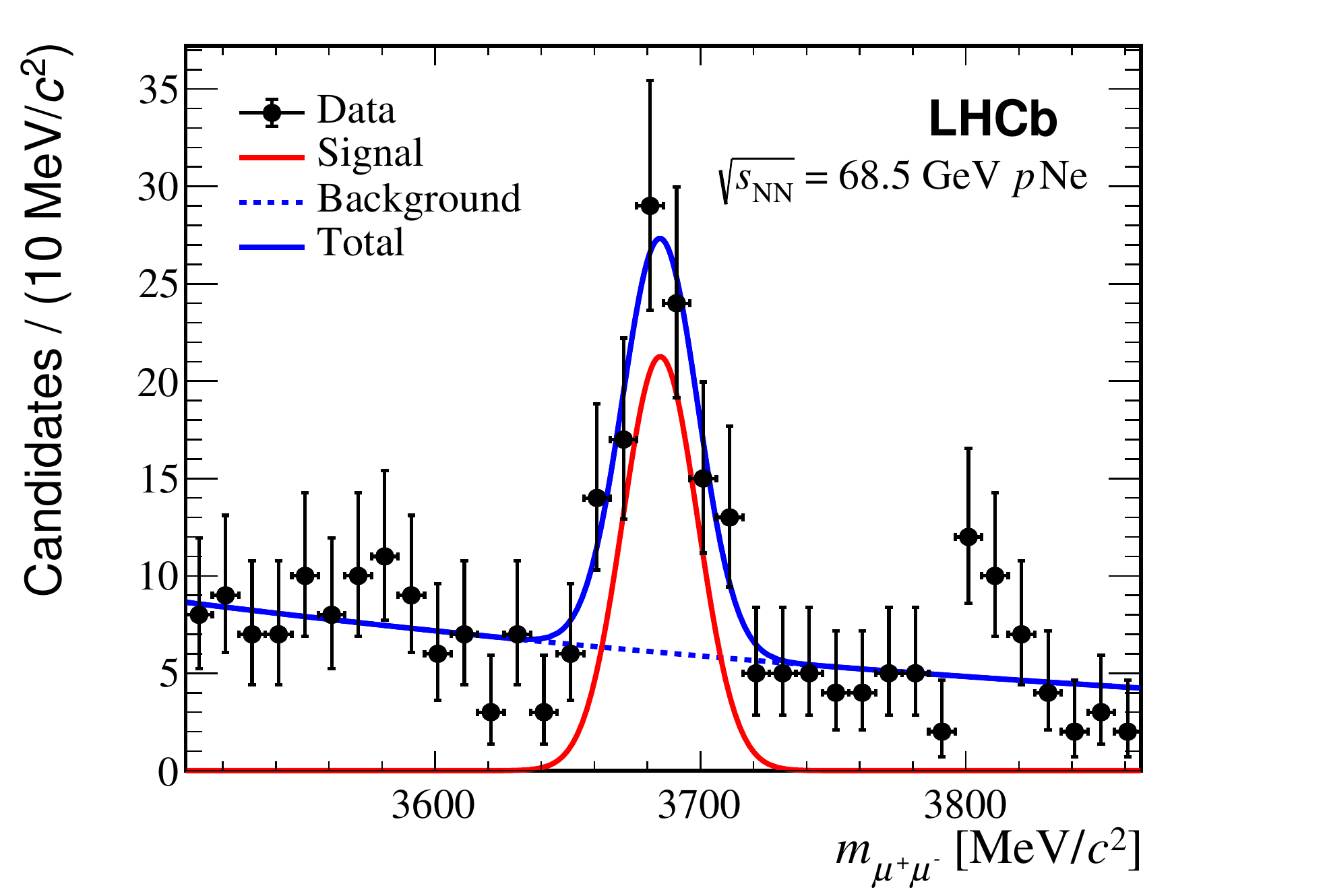}
\caption{Invariant mass distributions of (left) \mbox{$\jpsi\to\mun\mup$} candidates and (right) \mbox{$\psitwos\to\mun\mup$} candidates. The data are overlaid with the fit function.}
\label{fig::InvMass}
\end{center}
\end{figure}
%%%%% Offline selection
The offline selections of \jpsi and \psitwos candidates are similar to those used in Ref.~\cite{LHCb-PAPER-2018-023}. Events must contain a primary vertex with at least four tracks reconstructed in the VELO detector.  The \jpsi and \psitwos candidates are constructed from two oppositely-charged muons forming a good-quality vertex. The well-identified muons have a transverse momentum, \pt, larger than 500 MeV$/c$ and are required to be consistent with originating from the PV, which suppresses \jpsi and \psitwos mesons coming from b-hadron decays.
The measurements are performed in the ranges of transverse momentum $\pt < 8$~GeV$/c$ and rapidity $2.0 < y < 4.29$ of \jpsi and \psitwos mesons.
%%%%%% Efficiency
Corrections for the acceptance and reconstruction efficiencies are determined using samples of simulated  proton-neon (\pNe) collisions. 
In the simulation, \jpsi and \psitwos mesons are generated using \pythia 8 \cite{Pythia8} with a specific LHCb configuration \cite{1742-6596-331-3-032047}  and with colliding-proton beam momentum equal to the momentum per nucleon of the beam and target in the centre-of-mass frame. The decays are described by \evtgen \cite{EvtGen}, in which final-state radiation is generated using \photos \cite{PHOTOS}. The generated \jpsi and \psitwos meson decay products are embedded into \pNe minimum-bias events that are generated with the \epos event generator \cite{Pierog:2013ria}  using beam parameters obtained from data. Decays of hadrons generated with \epos are also described by \evtgen. The interaction of the generated particles with the detector and its response are implemented using the \geant toolkit \cite{geant2, geant4} as described in Ref. \cite{gauss}. After reconstruction, the simulated events are assigned weights, based on the VELO cluster multiplicity. This ensures that the event multiplicity and the PV position follow the same distributions as in the data.

Figure \ref{fig::InvMass} shows the invariant-mass distributions for the \jpsi and \psitwos candidates, from which the corresponding signal yields are obtained with extended maximum-likelihood fits, after all selection criteria are applied to the entire pNe data set. The signals are described by Crystal Ball functions \cite{CB} and the background shapes are modelled by exponential functions.  
The total  \jpsi and \psitwos signal yields are $4542\pm 71$ and $76\pm 12$, respectively. The signal yields are determined independently in intervals of \pt and $y^{\star}$. These yields are corrected for the total efficiencies, evaluated to 36.6$\%$ and 38.8$\%$ for the $J/\psi$ and $\psi(2S)$ respectively, which account for the geometrical acceptance of the detector, and the efficiencies of the trigger, event selection, PV and track reconstruction, and particle identification. Particle identification \cite{LHCb-PUB-2016-021} and tracking efficiencies are obtained from control samples in \pp collision data. All other efficiencies are determined using samples of simulated data. 

Several sources of systematic uncertainty are considered, affecting either the determination of the signal yields or the total efficiencies. They are summarised in Table~\ref{tab:systematics} separately
for contributions that are correlated and uncorrelated between different intervals of \pt and $y^\star$.
\begin{table}[t]
\caption{\small \label{tab:systematics} Systematic and statistical uncertainties on the \jpsi meson yield. Systematic uncertainties correlated between bins affect all measurements by the same relative amount. Ranges denote the minimum and the maximum values among the $y^{\star}$ or \pt intervals while the latter value is the uncertainty  integrated over $y^{\star}$ or \pt.}
\begin{center}
{\small
\begin{tabular}{@{}lc@{}}
\hline
{\textbf{Systematic uncertainties}} & \\
\hline
{\textbf{Uncorrelated between bins}} &  \\
Simulation sample size               & [1.4, \phantom{0}7.0]$\%$; 2.3 $\%$  \\ 
Signal determination                 & [1.4, 11.0]$\%$; 3.5 $\%$  \\
{\textbf{Correlated between bins}} & \\
Proton-proton collisions        &  2.0$\%$   \\
Neon purity & 1.2$\%$  \\
Tracking efficiency                & 1.1$\%$   \\
Particle identification efficiency & 1.1$\%$   \\
PV & 3.9$\%$  \\
Luminosity                          & 6.5$\%$ \\
\hline
\textbf{Statistical uncertainty}            & 1.6$\%$  \\
\hline
\end{tabular} 
}
\end{center}
\end{table}
Systematic uncertainty on the signal determination includes several contributions. A significant systematic uncertainty arises from the finite size of the simulation samples.
The systematic uncertainty associated to the determination of the signal yields is related to the mass fit. This uncertainty is evaluated using alternative models for signal and background shapes,  Gaussian and polynomial functions respectively, that reproduce the mass distributions equally well. The effect of the small (below $0.1\%$) residual contribution of signal from \bquark hadrons is investigated and found to be negligible.
Other contributions are obtained by determining the maximum contamination from residual \pp collisions with samples of pure \pNe collisions and pure \pp collisions.  The neon purity systematic uncertainty corresponds to the contamination from collisions between the beam and elements different from neon, coming from standard outgassing. It is quantified using data samples recorded with no neon injection. 
Since the tracking and particle identification efficiencies are determined using \pp control samples, the differences between the track multiplicity in \pNe and \pp collisions are considered as systematic uncertainties. The tracking and particle identification systematic uncertainties also take into account the size of the $pp$ control samples. 
The PV reconstruction systematic uncertainty corresponds to the variation of the efficiency over the whole $z_{PV}$ range, and to the difference between the PV reconstruction efficiency evaluated using the simulation and a data-driven approach exploiting the well-reconstructed $\phi\to K^+ K^-$ decay.
\begin{figure}[t]
\begin{center}
\includegraphics[width=0.9\textwidth]{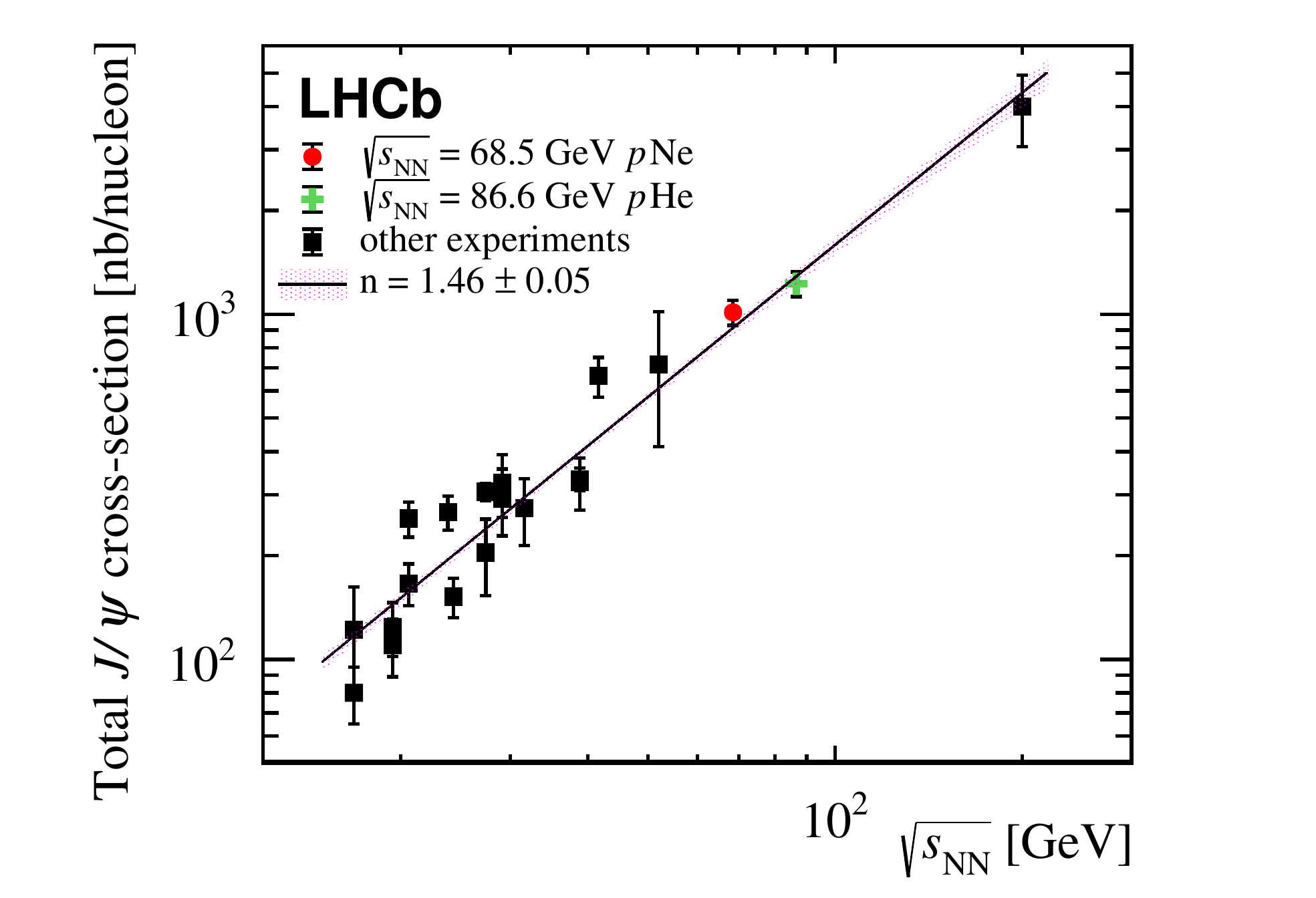}
\caption{Total \jpsi cross-section per target nucleon as a function of centre-of-mass energy. Experimental data, represented by black points, are taken from Ref. \cite{Maltoni:2006yp}. The red point corresponds to the \pNe result from the present analysis. The green point corresponds to a measurement performed by \lhcb with \pHe collisions \cite{LHCb-PAPER-2018-023}.\label{Fig:CompaXsection_Jpsi}}
\end{center}
\end{figure}
The integrated luminosity is determined to be $21.7 \pm $ 1.4 nb$^{-1}$ from the yield of electrons elastically scattering off the target Ne atoms as presented in Ref. \cite{antiproton}.
The measured \jpsi production cross-section per target nucleon and within $y^\star \in [-2.29, 0]$, using the world average branching fraction of $\jpsi \to \mu^+ \mu^-$ decays \cite{PDG2022}, is
\begin{equation*}
\sigma_{\jpsi} =  506 \pm 8 \pm  46  \textrm{ nb/nucleon},
\end{equation*}
where the first uncertainty is statistical and the second systematic.
To compare with previous experimental results at different energies, the \jpsi cross-section is extrapolated to the full phase space using \pythia 8 with the CT09MCS PDF set \cite{CT09MCS}, with no additional uncertainty related to the extrapolation, assuming forward-backward symmetry in the rapidity distribution. After extrapolation, the total \jpsi cross-section is
\begin{equation*}
\sigma_{\jpsi}^{4\pi} =  1013 \pm 16  \pm  92 \textrm{ nb/nucleon},
\end{equation*}
where the first uncertainty is statistical and the second systematic. An overview of \jpsi cross-section measurements performed at different centre-of-mass energies by different experiments \cite{Maltoni:2006yp}, including this measurement and the previous \lhcb  measurement in \pHe collisions at $\sqsnn=86.6\gev$  \cite{LHCb-PAPER-2018-023}, is shown in Fig. \ref{Fig:CompaXsection_Jpsi}. The data are well reproduced with the function $\sigma_{J/\psi}=C\times(\sqsnn)^n$, with $n=1.46\pm 0.05$ ($\chi^2/n_{\textrm{dof}}=64.6/18$ and $\textrm{p-value}=4\times 10^{-7}$), indicating a power-law dependence of the cross-section on the centre-of-mass energy, between $\sqsnn\sim20\gev$ and $\sqsnn=200\gev$.

\begin{figure}[t]
\begin{center}
\includegraphics[width=0.49\textwidth]{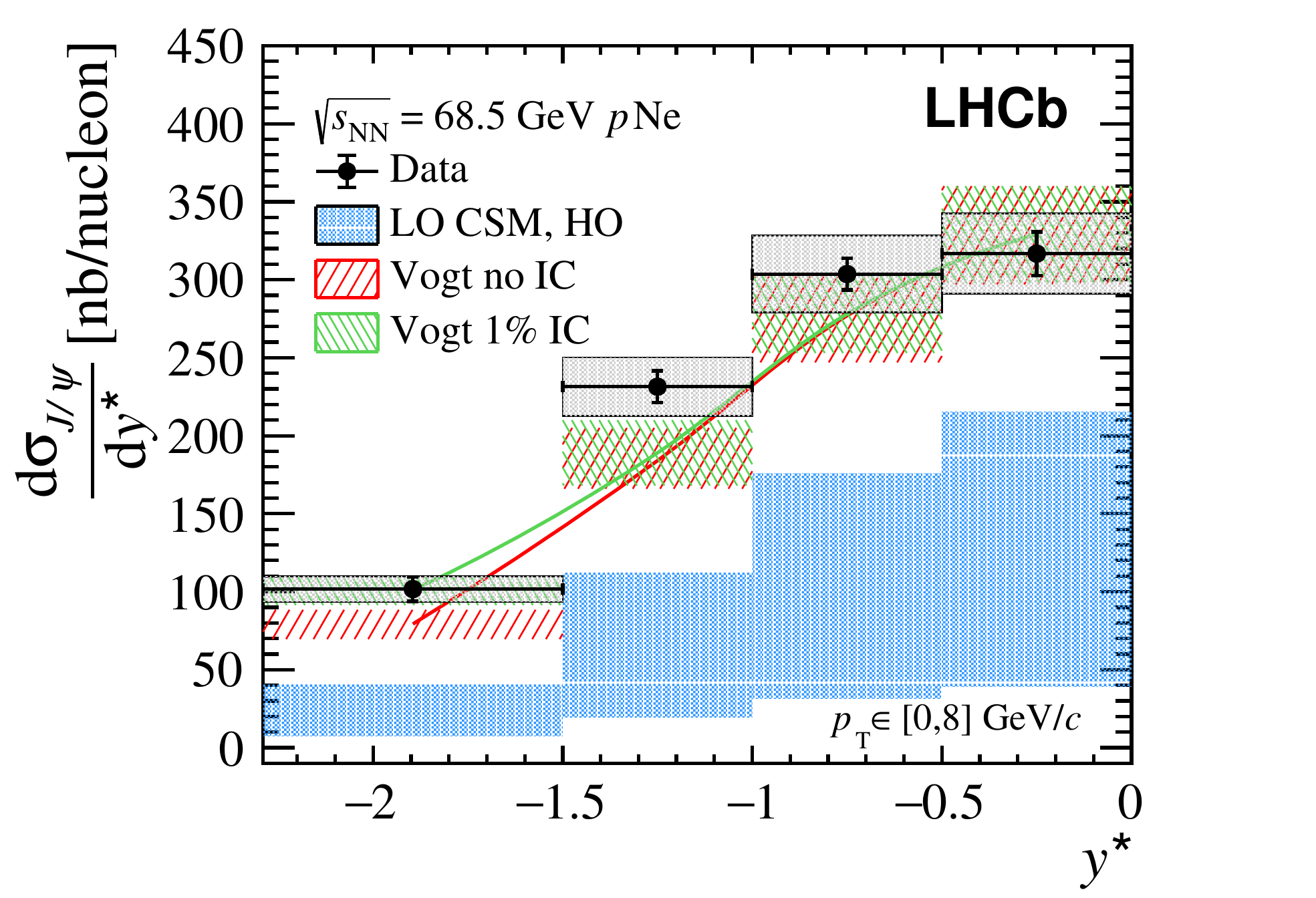}
\includegraphics[width=0.49\textwidth]{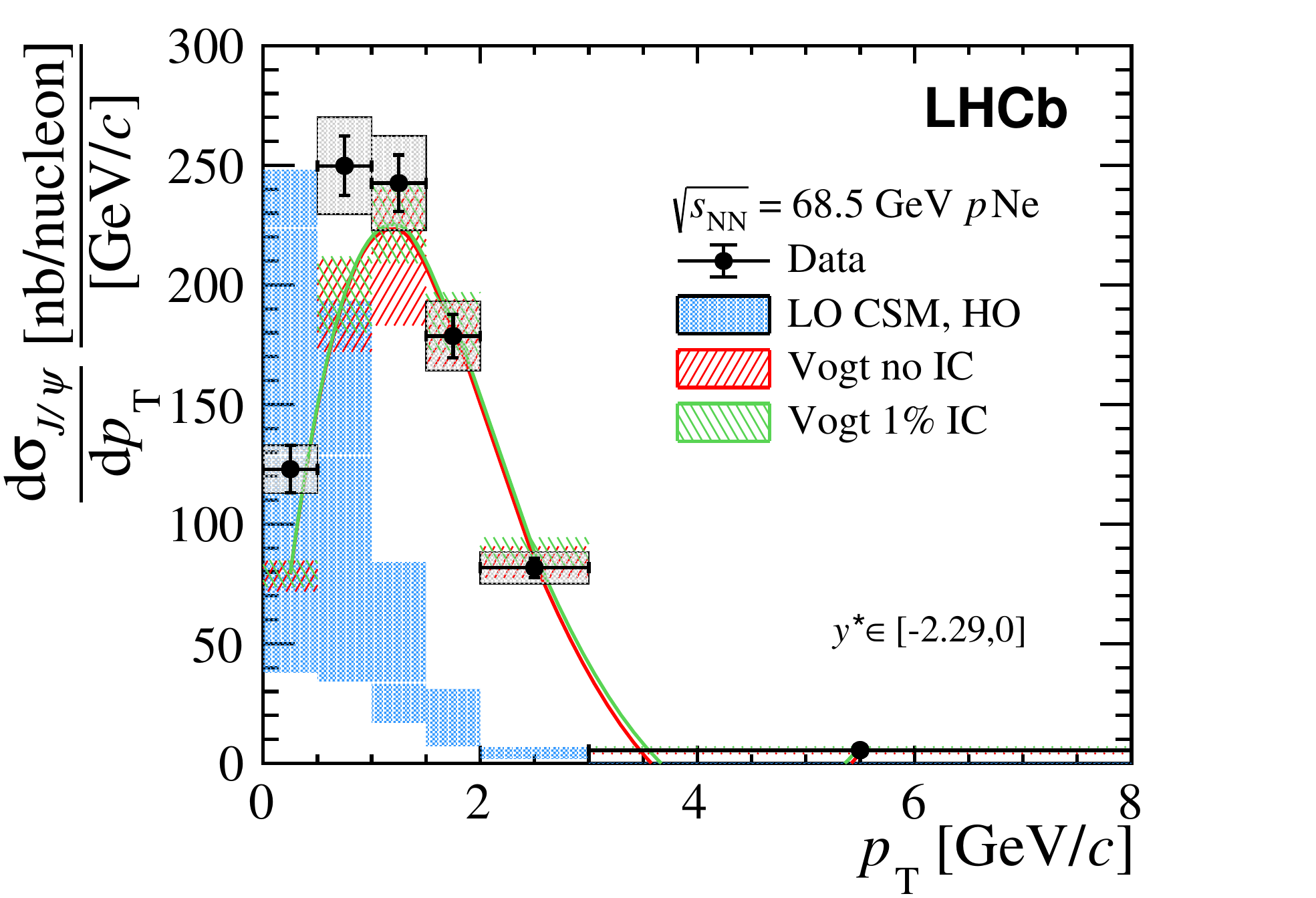}
\caption{Differential \jpsi cross-section as a function of (left) $y^\star$  and (right) \pt. The quadratic sums of statistical and uncorrelated systematic uncertainties are given by the error bars, while the grey boxes represent the correlated systematic uncertainties. Blue boxes (LO CSM, HO) correspond to predictions using the CT14NLO and nCTEQ15 PDF sets~\cite{Lansberg:2016deg,Shao:2015vga,Shao:2012iz,nCTEQ15}. Green and red boxes correspond to predictions (Vogt) from \cite{Vogt} with and without a 1$\%$ intrinsic charm (IC) contribution respectively (green and red lines indicate the central values). \label{Fig:Xsection_Jpsi_pT_y_average}}
\label{fig::cross-section}
\end{center}
\end{figure}

The \jpsi differential cross-sections per target nucleon, as functions of $y^\star$ and \pt, are shown in Fig.~\ref{Fig:Xsection_Jpsi_pT_y_average}. 
These results are compared with predictions of the HELAC-Onia (HO) generator~\cite{Lansberg:2016deg,Shao:2015vga,Shao:2012iz}, using QCD Leading Order (LO) calculations within the Color Singlet Model (CSM), with the proton CT14NLO and nuclear nCTEQ15 PDF~\cite{nCTEQ15} sets. The error band is obtained by varying the renormalization and factorization scales from 0.5 to 2.
These predictions underestimate the measured total cross-sections. The data are better described by alternative predictions (Vogt), using calculations in the Color Evaporation Model carried out at Next-to-Leading Order (NLO) in the heavy-flavour cross-section, with or without a 1$\%$ intrinsic charm (IC) contribution \cite{Vogt}. 

\begin{figure}[t]
\begin{center}
\includegraphics[width=0.49\textwidth]{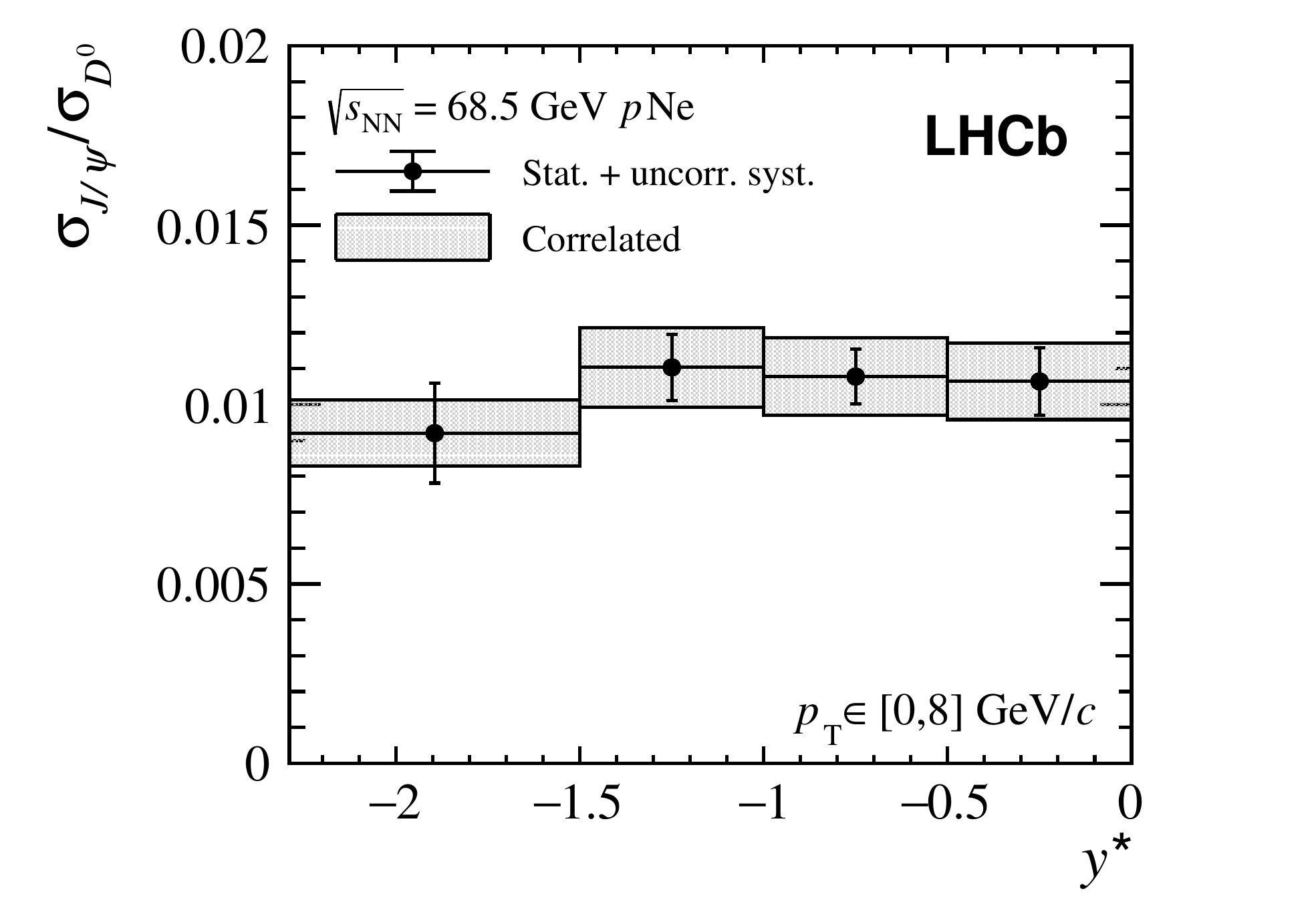}
\includegraphics[width=0.49\textwidth]{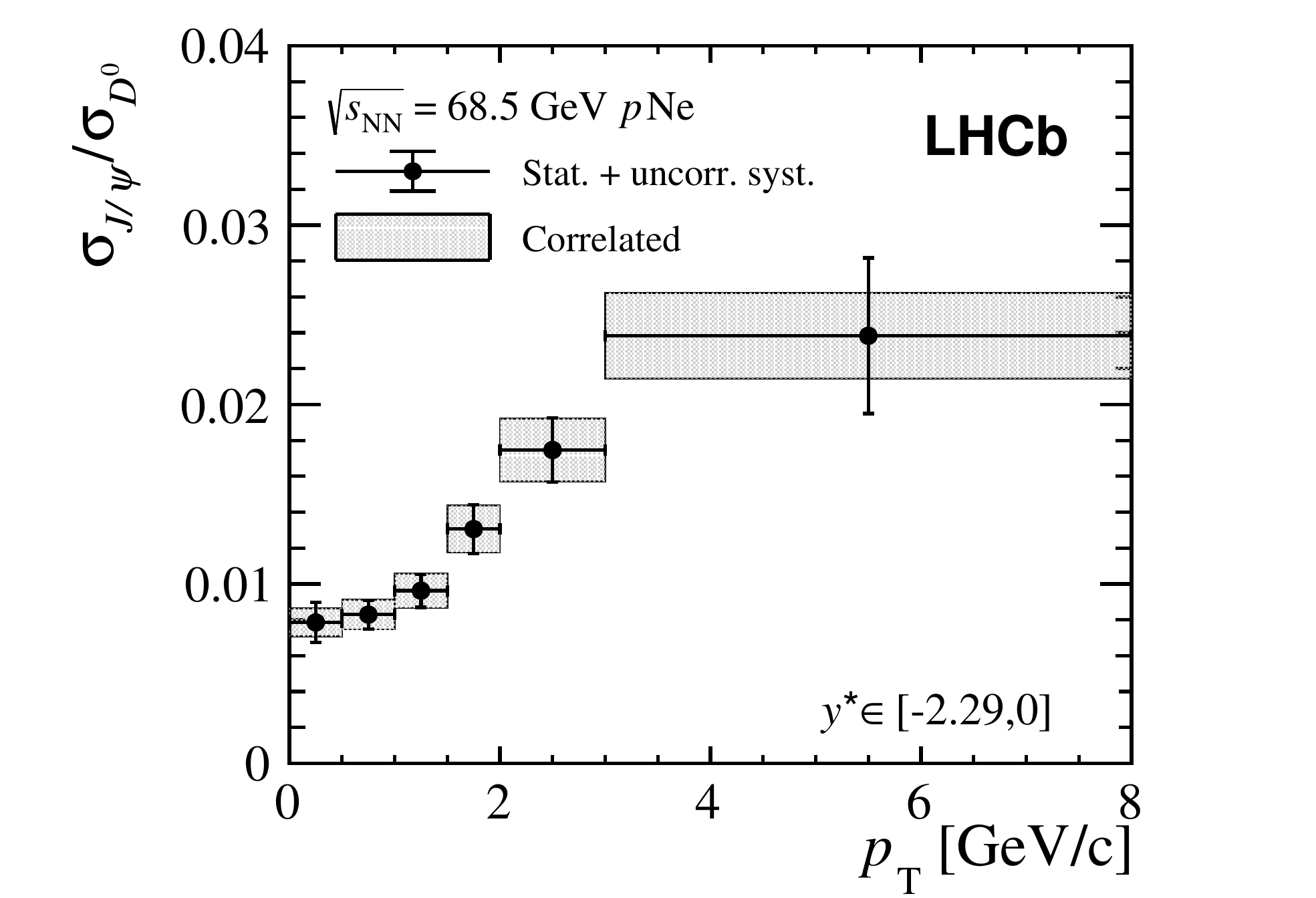}
\caption{Ratio of \jpsi and  \Dz cross-sections  as a function of (left) $y^\star$  and (right) \pt. The quadratic sums of the statistical and uncorrelated systematic uncertainties are given by the error bars, while the grey boxes represent the correlated systematic uncertainties.}
\label{fig::psiD0}
\end{center}
\end{figure}

The \jpsi production cross-section is also compared to the \Dz production cross-section extracted from the same dataset, in the same kinematical conditions \cite{D0pNe}. Several systematic uncertainties cancel in the $\jpsi/\Dz$ cross-section ratio, related to the PV and track reconstruction efficiencies, the contamination from residual \pp collisions,  the neon purity and the luminosity determination. The ratio of  \jpsi and \Dz cross-sections is

\begin{equation*}
    \frac{\sigma_\jpsi}{\sigma_\Dz} = (1.06\pm0.02\pm 0.09) \%,
\end{equation*}
where the first uncertainty is statistical and the second systematic. The ratio takes into account the branching fractions 
\cite{PDG2022}
of $\jpsi\to\mup\mun$ and $\Dz\to\Km\pip$.
The \jpsi-to-\Dz cross-section ratio as a function of $y^\star$ and \pt is shown in Fig. \ref{fig::psiD0}. Although this ratio shows a strong dependence on \pt, the data show no significant rapidity dependence.

\begin{figure}
\begin{center}
\includegraphics[width=0.80\textwidth]{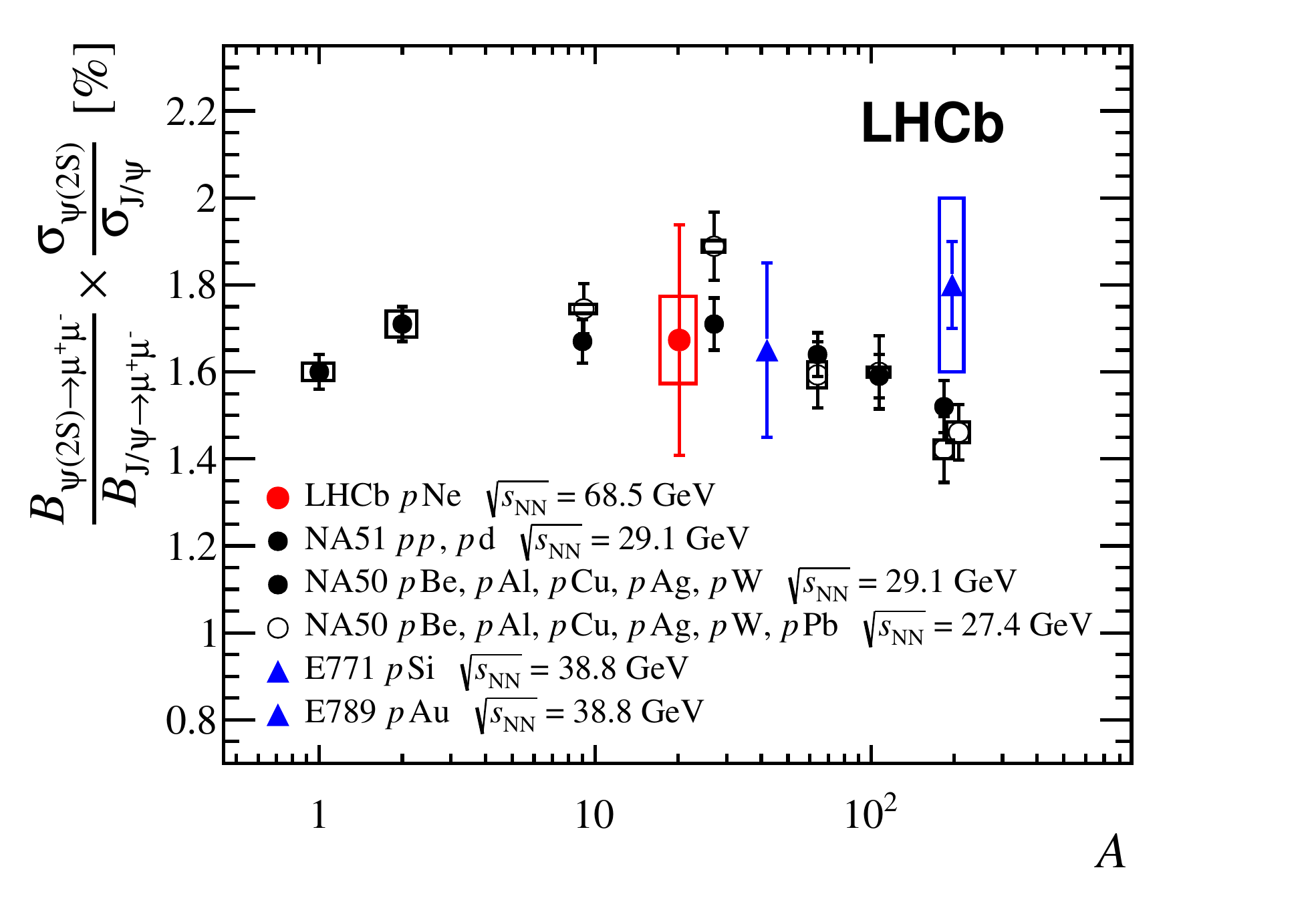} 
\caption{The \psitwos-to-\jpsi production ratio as a function of the target atomic mass number $A$. The red point corresponds to the $\sqsnn=68.5\gev$ \pNe result from the present analysis, vertical error bar corresponds to the statistical uncertainty and the box to the systematic uncertainty. The other points show previous fixed-target experimental data at various centre-of-mass energies \cite{NA51PLB438,NA50EPJC33,NA50EPJC48,E771PLB374,E789PRD52}.}
\label{fig::psiPsiprim}
\end{center}
\end{figure}

The \psitwos production cross-section is also measured.
Due to the limited size of the \psitwos sample, only the relative production rate of \psitwos and \jpsi mesons is presented, where most of the efficiencies and systematic uncertainties cancel out. The remaining systematic uncertainties are evaluated to be $0.01\%$ for the finite size of the simulation sample, $0.09\%$ for the total efficiency differences between \jpsi and \psitwos and  0.05$\%$ for the signal extraction. The relative production rate of \psitwos and \jpsi mesons is
\begin{equation*}
    \frac{\BF_{\psitwos\to\mup\mun}}{\BF_{\jpsi\to\mup\mun}}\times\frac{\sigma_\psitwos}{\sigma_\jpsi} = (1.67 \pm 0.27\pm 0.10)\%,
\end{equation*}
where $\BF_{\psitwos\to\mup\mun}$ and $\BF_{\jpsi\to\mup\mun}$ are the branching fractions of $\psitwos\to\mup\mun$ and $\jpsi\to\mup\mun$ decays, respectively, and  the first uncertainty is statistical and the second systematic. Figure \ref{fig::psiPsiprim} compares this result to measurements performed at various centre-of-mass energies by other experiments as a function of the target atomic mass number $A$ \cite{NA51PLB438,NA50EPJC33,NA50EPJC48,E771PLB374,E789PRD52}. Measurement is in agreement with other proton-nucleus measurements at similar values of $A$.

In summary, the study of charmonium production in \pNe collisions at $\sqsnn=68.5\gev$  recorded by the \lhcb experiment is presented. The \jpsi production cross-section is measured in the centre-of-mass rapidity range $y^\star\in[-2.29,0]$. The comparison of this new measurement with earlier data supports a power-law dependence of the \jpsi production cross-section on centre-of-mass energy.
The \jpsi-to-\Dz cross-section ratio is found to be independent  of rapidity and the \psitwos-to-\jpsi cross-section ratio is found to be
$(1.67 \pm 0.27\pm 0.10)\%$. This result is
in a good agreement with other measurements involving beam and target nuclei of similar sizes, and performed at different centre-of-mass energies. 

\section*{Acknowledgements}
%
% These Acknowledgements valid from 3-May-2019
%
\noindent We express our gratitude to our colleagues in the CERN
accelerator departments for the excellent performance of the LHC. We
thank the technical and administrative staff at the LHCb
institutes.
We acknowledge support from CERN and from the national agencies:
CAPES, CNPq, FAPERJ and FINEP (Brazil); 
MOST and NSFC (China); 
CNRS/IN2P3 (France); 
BMBF, DFG and MPG (Germany); 
INFN (Italy); 
NWO (Netherlands); 
MNiSW and NCN (Poland); 
MEN/IFA (Romania); 
%MSHE (Russia); 
MICINN (Spain); 
SNSF and SER (Switzerland); 
NASU (Ukraine); 
STFC (United Kingdom); 
DOE NP and NSF (USA).
We acknowledge the computing resources that are provided by CERN, IN2P3
(France), KIT and DESY (Germany), INFN (Italy), SURF (Netherlands),
PIC (Spain), GridPP (United Kingdom), 
%RRCKI and Yandex LLC (Russia), 
CSCS (Switzerland), IFIN-HH (Romania), CBPF (Brazil),
Polish WLCG  (Poland) and NERSC (USA).
We are indebted to the communities behind the multiple open-source
software packages on which we depend.
Individual groups or members have received support from
ARC and ARDC (Australia);
Minciencias (Colombia);
AvH Foundation (Germany);
EPLANET, Marie Sk\l{}odowska-Curie Actions and ERC (European Union);
A*MIDEX, ANR, IPhU and GLUODYNAMICS/Labex P2IO, and R\'{e}gion Auvergne-Rh\^{o}ne-Alpes (France);
Key Research Program of Frontier Sciences of CAS, CAS PIFI, CAS CCEPP, 
Fundamental Research Funds for the Central Universities, 
and Sci. \& Tech. Program of Guangzhou (China);
%Key Research Program of Frontier Sciences of CAS, CAS PIFI,
%Thousand Talents Program, and Sci. \& Tech. Program of Guangzhou (China);
%RFBR, RSF and Yandex LLC (Russia);
GVA, XuntaGal, GENCAT and Prog.~Atracci\'on Talento, CM (Spain);
SRC (Sweden);
the Leverhulme Trust, the Royal Society
 and UKRI (United Kingdom).

\clearpage 

%%%%% bibliography
\addcontentsline{toc}{section}{References}
\setboolean{inbibliography}{true}
%\bibliographystyle{LHCb}
%%%%%%%%%%%%%%%%%%%%%%%%%%%
\ifx\mcitethebibliography\mciteundefinedmacro
\PackageError{LHCb.bst}{mciteplus.sty has not been loaded}
{This bibstyle requires the use of the mciteplus package.}\fi
\providecommand{\href}[2]{#2}

\clearpage
% LHCb collaboration author list
% Data extracted on May 12th, 2023 at 3:10pm for paper reference LHCb-PAPER-2022-014
\centerline
{\large\bf LHCb collaboration}
\begin
{flushleft}
\small
R.~Aaij$^{32}$\lhcborcid{0000-0003-0533-1952},
A.S.W.~Abdelmotteleb$^{50}$\lhcborcid{0000-0001-7905-0542},
C.~Abellan~Beteta$^{44}$,
F.~Abudin{\'e}n$^{50}$\lhcborcid{0000-0002-6737-3528},
T.~Ackernley$^{54}$\lhcborcid{0000-0002-5951-3498},
B.~Adeva$^{40}$\lhcborcid{0000-0001-9756-3712},
M.~Adinolfi$^{48}$\lhcborcid{0000-0002-1326-1264},
H.~Afsharnia$^{9}$,
C.~Agapopoulou$^{13}$\lhcborcid{0000-0002-2368-0147},
C.A.~Aidala$^{76}$\lhcborcid{0000-0001-9540-4988}
% ...

{\footnotesize \it

% ...
$^{9}$Universit{\'e} Clermont Auvergne, CNRS/IN2P3, LPC, Clermont-Ferrand, France\\
% ...
$^{13}$LPNHE, Sorbonne Universit{\'e}, Paris Diderot Sorbonne Paris Cit{\'e}, CNRS/IN2P3, Paris, France\\
% ...
$^{32}$Nikhef National Institute for Subatomic Physics, Amsterdam, Netherlands\\
% ...
$^{40}$Instituto Galego de F{\'\i}sica de Altas Enerx{\'\i}as (IGFAE), Universidade de Santiago de Compostela, Santiago de Compostela, Spain\\
% ...
$^{44}$Physik-Institut, Universit{\"a}t Z{\"u}rich, Z{\"u}rich, Switzerland\\
% ...
$^{48}$H.H. Wills Physics Laboratory, University of Bristol, Bristol, United Kingdom\\
% ...
$^{50}$Department of Physics, University of Warwick, Coventry, United Kingdom\\
% ...
$^{54}$Oliver Lodge Laboratory, University of Liverpool, Liverpool, United Kingdom\\
% ...
$^{62}$Syracuse University, Syracuse, NY, United States\\
% ...
$^{76}$University of Michigan, Ann Arbor, MI, United States, associated to $^{62}$\\
\bigskip
% ...
}
\end{flushleft}


\begin{mcitethebibliography}{10}
\mciteSetBstSublistMode{n}
\mciteSetBstMaxWidthForm{subitem}{\alph{mcitesubitemcount})}
\mciteSetBstSublistLabelBeginEnd{\mcitemaxwidthsubitemform\space}
{\relax}{\relax}

\bibitem{NAPLB207}
C.~Gerschel and J.~H\"ufner, \ifthenelse{\boolean{articletitles}}{\emph{A
  contribution to the suppression of the \jpsi meson produced in high-energy
  nucleus-nucleus collisions},
  }{}\href{http://dx.doi.org/10.1016/0370-2693(88)90570-9}{Phys.\ Lett.\
  \textbf{B207} (1988) 253}\relax
\mciteBstWouldAddEndPuncttrue
\mciteSetBstMidEndSepPunct{\mcitedefaultmidpunct}
{\mcitedefaultendpunct}{\mcitedefaultseppunct}\relax
\EndOfBibitem
\bibitem{PRD77}
Z.-B.~Kang and J.-W.~Qiu, \ifthenelse{\boolean{articletitles}}{\emph{Transverse momentum broadening of vector boson production in high energy nuclear collisions},
  }{}\href{http://dx.doi.org/10.1103/PhysRevD.77.114027}{Phys.\ Rev.\
  \textbf{D77} (2008) 114027}\relax
\mciteBstWouldAddEndPuncttrue
\mciteSetBstMidEndSepPunct{\mcitedefaultmidpunct}
{\mcitedefaultendpunct}{\mcitedefaultseppunct}\relax
\EndOfBibitem
\bibitem{ELJHEP03}
F.~Arleo and S.~Peign\'e,
  \ifthenelse{\boolean{articletitles}}{\emph{Heavy-quarkonium suppression in
  p-A collisions from parton energy loss in cold qcd matter},
  }{}\href{http://dx.doi.org/10.1007/JHEP03(2013)122}{JHEP \textbf{03} (2013)
  122}, \href{http://arxiv.org/abs/1212.0434}{{\normalfont\ttfamily
  arXiv:1212.0434}}\relax
\mciteBstWouldAddEndPuncttrue
\mciteSetBstMidEndSepPunct{\mcitedefaultmidpunct}
{\mcitedefaultendpunct}{\mcitedefaultseppunct}\relax
\EndOfBibitem
\bibitem{CapellaPRL85}
A.~Capella {\em et~al.},
  \ifthenelse{\boolean{articletitles}}{\emph{Nonsaturation of the \jpsi
  suppression at large transverse energy in the comovers approach},
  }{}\href{http://dx.doi.org/10.1103/PhysRevLett.85.2080}{Phys.\ Rev.\ Lett.\
  \textbf{85} (2000) 2080},
  \href{http://arxiv.org/abs/hep-ph/0002300}{{\normalfont\ttfamily
  arXiv:hep-ph/0002300}}\relax
\mciteBstWouldAddEndPuncttrue
\mciteSetBstMidEndSepPunct{\mcitedefaultmidpunct}
{\mcitedefaultendpunct}{\mcitedefaultseppunct}\relax
\EndOfBibitem
\bibitem{PDFEPJC77}
K.~J. Eskola {\em et~al.}, \ifthenelse{\boolean{articletitles}}{\emph{Epps16:
  nuclear parton distributions with {LHC} data},
  }{}\href{http://dx.doi.org/10.1140/epjc/s10052-017-4725-9}{Eur.\ Phys.\ J.\
  \textbf{C77} (2017) 163},
  \href{http://arxiv.org/abs/1612.05741}{{\normalfont\ttfamily
  arXiv:1612.05741}}\relax
\mciteBstWouldAddEndPuncttrue
\mciteSetBstMidEndSepPunct{\mcitedefaultmidpunct}
{\mcitedefaultendpunct}{\mcitedefaultseppunct}\relax
\EndOfBibitem
\bibitem{PDFPRD93}
K.~Kovarik {\em et~al.}, \ifthenelse{\boolean{articletitles}}{\emph{ncteq15:
  Global analysis of nuclear parton distributions with uncertainties in the
  cteq framework},
  }{}\href{http://dx.doi.org/10.1103/PhysRevD.93.085037}{Phys.\ Rev.\
  \textbf{D93} (2016) 085037},
  \href{http://arxiv.org/abs/1509.00792}{{\normalfont\ttfamily
  arXiv:1509.00792}}\relax
\mciteBstWouldAddEndPuncttrue
\mciteSetBstMidEndSepPunct{\mcitedefaultmidpunct}
{\mcitedefaultendpunct}{\mcitedefaultseppunct}\relax
\EndOfBibitem
\bibitem{D0pNe}
LHCb collaboration, R.~Aaij {\em et~al.},
  \ifthenelse{\boolean{articletitles}}{\emph{Open charm production and
  asymmetry in {$p$Ne} collisions at $\sqsnn=68.5 \gev$}, }{}
  \href{https://arxiv.org/abs/2211.11633}{{\normalfont\ttfamily
  arXiv:2211.11633}}\relax
\mciteBstWouldAddEndPuncttrue
\mciteSetBstMidEndSepPunct{\mcitedefaultmidpunct}
{\mcitedefaultendpunct}{\mcitedefaultseppunct}\relax
\EndOfBibitem
\bibitem{Alves:2008zz}
LHCb collaboration, A.~A. Alves~Jr.\ {\em et~al.},
  \ifthenelse{\boolean{articletitles}}{\emph{{The \lhcb detector at the LHC}},
  }{}\href{http://dx.doi.org/10.1088/1748-0221/3/08/S08005}{JINST \textbf{3}
  (2008) S08005}\relax
\mciteBstWouldAddEndPuncttrue
\mciteSetBstMidEndSepPunct{\mcitedefaultmidpunct}
{\mcitedefaultendpunct}{\mcitedefaultseppunct}\relax
\EndOfBibitem
\bibitem{LHCb-DP-2014-002}
LHCb collaboration, R.~Aaij {\em et~al.},
  \ifthenelse{\boolean{articletitles}}{\emph{{LHCb detector performance}},
  }{}\href{http://dx.doi.org/10.1142/S0217751X15300227}{Int.\ J.\ Mod.\ Phys.\
  \textbf{A30} (2015) 1530022},
  \href{http://arxiv.org/abs/1412.6352}{{\normalfont\ttfamily
  arXiv:1412.6352}}\relax
\mciteBstWouldAddEndPuncttrue
\mciteSetBstMidEndSepPunct{\mcitedefaultmidpunct}
{\mcitedefaultendpunct}{\mcitedefaultseppunct}\relax
\EndOfBibitem
\bibitem{LHCb-PAPER-2014-047}
LHCb collaboration, R.~Aaij {\em et~al.},
  \ifthenelse{\boolean{articletitles}}{\emph{{Precision luminosity measurements
  at LHCb}}, }{}\href{http://dx.doi.org/10.1088/1748-0221/9/12/P12005}{JINST
  \textbf{9} (2014) P12005},
  \href{http://arxiv.org/abs/1410.0149}{{\normalfont\ttfamily
  arXiv:1410.0149}}\relax
\mciteBstWouldAddEndPuncttrue
\mciteSetBstMidEndSepPunct{\mcitedefaultmidpunct}
{\mcitedefaultendpunct}{\mcitedefaultseppunct}\relax
\EndOfBibitem
\bibitem{Barschel}
C.~Barschel,
  \ifthenelse{\boolean{articletitles}}{\emph{{Precision luminosity measurements at LHCb with beam-gas imaging}}, PhD thesis, RWTH Aachen University, 2014, 
  }{}\href{https://cds.cern.ch/record/1693671}{CERN-THESIS-2013-301}\relax
\mciteBstWouldAddEndPuncttrue
\mciteSetBstMidEndSepPunct{\mcitedefaultmidpunct}
{\mcitedefaultendpunct}{\mcitedefaultseppunct}\relax
\EndOfBibitem
\bibitem{trigger}
LHCb collaboration, R.~Aaij {\em et~al.},
  \ifthenelse{\boolean{articletitles}}{\emph{{The LHCb trigger and its
  performance in 2011}},
  }{}\href{http://dx.doi.org/10.1088/1748-0221/8/04/P04022}{JINST \textbf{8}
  (2013) P04022}, \href{http://arxiv.org/abs/1211.3055}{{\normalfont\ttfamily
  arXiv:1211.3055}}\relax
\mciteBstWouldAddEndPuncttrue
\mciteSetBstMidEndSepPunct{\mcitedefaultmidpunct}
{\mcitedefaultendpunct}{\mcitedefaultseppunct}\relax
\EndOfBibitem
\bibitem{LHCb-PAPER-2018-023}
LHCb collaboration, R.~Aaij {\em et~al.},
  \ifthenelse{\boolean{articletitles}}{\emph{{First measurement of charm
  production in fixed-target configuration at the LHC}},
  }{}\href{http://dx.doi.org/10.1103/PhysRevLett.122.132002}{Phys.\ Rev.\
  Lett.\  \textbf{122} (2019) 132002},
  \href{http://arxiv.org/abs/1810.07907}{{\normalfont\ttfamily
  arXiv:1810.07907}}\relax
\mciteBstWouldAddEndPuncttrue
\mciteSetBstMidEndSepPunct{\mcitedefaultmidpunct}
{\mcitedefaultendpunct}{\mcitedefaultseppunct}\relax
\EndOfBibitem
\bibitem{Pythia8}
T.~Sj{\"o}strand, S.~Mrenna, and P.~Skands,
  \ifthenelse{\boolean{articletitles}}{\emph{{A brief introduction to \pythia
  8.1}}, }{}\href{http://dx.doi.org/10.1016/j.cpc.2008.01.036}{Comput.\ Phys.\
  Commun.\  \textbf{178} (2008) 852},
  \href{http://arxiv.org/abs/0710.3820}{{\normalfont\ttfamily
  arXiv:0710.3820}}\relax
\mciteBstWouldAddEndPuncttrue
\mciteSetBstMidEndSepPunct{\mcitedefaultmidpunct}
{\mcitedefaultendpunct}{\mcitedefaultseppunct}\relax
\EndOfBibitem
\bibitem{1742-6596-331-3-032047}
LHCb collaboration, I.~Belyaev {\em et~al.},
  \ifthenelse{\boolean{articletitles}}{\emph{{Handling of the generation of
  primary events in Gauss, the LHCb simulation framework}},
  }{}\href{http://dx.doi.org/10.1088/1742-6596/331/3/032047}{J.\ Phys.\ Conf.\
  Ser.\  \textbf{331} (2011) 032047}\relax
\mciteBstWouldAddEndPuncttrue
\mciteSetBstMidEndSepPunct{\mcitedefaultmidpunct}
{\mcitedefaultendpunct}{\mcitedefaultseppunct}\relax
\EndOfBibitem
\bibitem{EvtGen}
D.~J. Lange, \ifthenelse{\boolean{articletitles}}{\emph{{The EvtGen particle
  decay simulation package}},
  }{}\href{http://dx.doi.org/10.1016/S0168-9002(01)00089-4}{Nucl.\ Instrum.\
  Meth.\  \textbf{A462} (2001) 152}\relax
\mciteBstWouldAddEndPuncttrue
\mciteSetBstMidEndSepPunct{\mcitedefaultmidpunct}
{\mcitedefaultendpunct}{\mcitedefaultseppunct}\relax
\EndOfBibitem
\bibitem{PHOTOS}
P.~Golonka and Z.~Was, \ifthenelse{\boolean{articletitles}}{\emph{{PHOTOS Monte
  Carlo: a precision tool for QED corrections in $Z$ and $W$ decays}},
  }{}\href{http://dx.doi.org/10.1140/epjc/s2005-02396-4}{Eur.\ Phys.\ J.\
  \textbf{C45} (2006) 97},
  \href{http://arxiv.org/abs/hep-ph/0506026}{{\normalfont\ttfamily
  arXiv:hep-ph/0506026}}\relax
\mciteBstWouldAddEndPuncttrue
\mciteSetBstMidEndSepPunct{\mcitedefaultmidpunct}
{\mcitedefaultendpunct}{\mcitedefaultseppunct}\relax
\EndOfBibitem
\bibitem{Pierog:2013ria}
T.~Pierog {\em et~al.}, \ifthenelse{\boolean{articletitles}}{\emph{{EPOS LHC:
  Test of collective hadronization with data measured at the CERN Large Hadron
  Collider}}, }{}\href{http://dx.doi.org/10.1103/PhysRevC.92.034906}{Phys.\
  Rev.\  \textbf{C92} (2015) 034906},
  \href{http://arxiv.org/abs/1306.0121}{{\normalfont\ttfamily
  arXiv:1306.0121}}\relax
\mciteBstWouldAddEndPuncttrue
\mciteSetBstMidEndSepPunct{\mcitedefaultmidpunct}
{\mcitedefaultendpunct}{\mcitedefaultseppunct}\relax
\EndOfBibitem
\bibitem{geant2}
\geant collaboration, J.~Allison {\em et~al.},
  \ifthenelse{\boolean{articletitles}}{\emph{\geant developments and
  applications}, }{}\href{http://dx.doi.org/10.1109/TNS.2006.869826}{IEEE
  Trans.\ Nucl.\ Sci.\  \textbf{53} (2006) 270}\relax
\mciteBstWouldAddEndPuncttrue
\mciteSetBstMidEndSepPunct{\mcitedefaultmidpunct}
{\mcitedefaultendpunct}{\mcitedefaultseppunct}\relax
\EndOfBibitem
\bibitem{geant4}
\geant collaboration, S.~Agostinelli {\em et~al.},
  \ifthenelse{\boolean{articletitles}}{\emph{\geant: a simulation toolkit},
  }{}\href{http://dx.doi.org/10.1016/S0168-9002(03)01368-8}{Nucl.\ Instr.\
  Meth.\  \textbf{506} (2003) 250}\relax
\mciteBstWouldAddEndPuncttrue
\mciteSetBstMidEndSepPunct{\mcitedefaultmidpunct}
{\mcitedefaultendpunct}{\mcitedefaultseppunct}\relax
\EndOfBibitem
\bibitem{gauss}
M.~Clemencic {\em et~al.}, \ifthenelse{\boolean{articletitles}}{\emph{{The LHCb
  simulation application, Gauss: Design, evolution and experience}},
  }{}\href{http://dx.doi.org/10.1088/1742-6596/331/3/032023}{J.\ Phys.\ Conf.\
  Ser.\  \textbf{331} (2011) 032023}\relax
\mciteBstWouldAddEndPuncttrue
\mciteSetBstMidEndSepPunct{\mcitedefaultmidpunct}
{\mcitedefaultendpunct}{\mcitedefaultseppunct}\relax
\EndOfBibitem
\bibitem{CB}
T.~Skwarnicki, {\em {A study of the radiative cascade transitions between the
  Upsilon-Prime and Upsilon resonances}}, PhD thesis, Institute of Nuclear
  Physics, Krakow, 1986, DESY-F31-86-02\relax
\mciteBstWouldAddEndPuncttrue
\mciteSetBstMidEndSepPunct{\mcitedefaultmidpunct}
{\mcitedefaultendpunct}{\mcitedefaultseppunct}\relax
\EndOfBibitem
\bibitem{LHCb-PUB-2016-021}
L.~Anderlini {\em et~al.}, \ifthenelse{\boolean{articletitles}}{\emph{{The
  PIDCalib package}}, }{}
  \href{http://cdsweb.cern.ch/search?p=LHCb-PUB-2016-021&f=reportnumber&action_search=Search&c=LHCb+Notes}
  {LHCb-PUB-2016-021}\relax
\mciteBstWouldAddEndPuncttrue
\mciteSetBstMidEndSepPunct{\mcitedefaultmidpunct}
{\mcitedefaultendpunct}{\mcitedefaultseppunct}\relax
\EndOfBibitem
\bibitem{Maltoni:2006yp}
F.~Maltoni {\em et~al.}, \ifthenelse{\boolean{articletitles}}{\emph{{Analysis
  of charmonium production at fixed-target experiments in the NRQCD approach}},
  }{}\href{http://dx.doi.org/10.1016/j.physletb.2006.05.010}{Phys.\ Lett.\
  \textbf{B638} (2006) 202},
  \href{http://arxiv.org/abs/hep-ph/0601203}{{\normalfont\ttfamily
  arXiv:hep-ph/0601203}}\relax
\mciteBstWouldAddEndPuncttrue
\mciteSetBstMidEndSepPunct{\mcitedefaultmidpunct}
{\mcitedefaultendpunct}{\mcitedefaultseppunct}\relax
\EndOfBibitem
\bibitem{antiproton}
LHCb collaboration, R.~Aaij {\em et~al.},
  \ifthenelse{\boolean{articletitles}}{\emph{{Measurement of antiproton
  production in {$p$He} collisions at $\sqsnn=110$ GeV}},
  }{}\href{http://dx.doi.org/10.1103/PhysRevLett.121.222001}{Phys.\ Rev.\
  Lett.\  \textbf{121} (2018) 222001},
  \href{http://arxiv.org/abs/1808.06127}{{\normalfont\ttfamily
  arXiv:1808.06127}}\relax
\mciteBstWouldAddEndPuncttrue
\mciteSetBstMidEndSepPunct{\mcitedefaultmidpunct}
{\mcitedefaultendpunct}{\mcitedefaultseppunct}\relax
\EndOfBibitem
\bibitem{PDG2022}
Particle Data Group, R.~L. Workman {\em et~al.},
  \ifthenelse{\boolean{articletitles}}{\emph{{\href{http://pdg.lbl.gov/}{Review
  of particle physics}}},
  }{}\href{http://dx.doi.org/10.1093/ptep/ptac097}{Prog.\ Theor.\ Exp.\ Phys.\
  \textbf{2022} (2022), no.~8 083C01}\relax
\mciteBstWouldAddEndPuncttrue
\mciteSetBstMidEndSepPunct{\mcitedefaultmidpunct}
{\mcitedefaultendpunct}{\mcitedefaultseppunct}\relax
\EndOfBibitem
\bibitem{CT09MCS}
H.-L. Lai {\em et~al.}, \ifthenelse{\boolean{articletitles}}{\emph{{Parton
  distributions for event generators}},
  }{}\href{http://dx.doi.org/10.1007/JHEP04(2010)035}{JHEP \textbf{04} (2010)
  035}, \href{http://arxiv.org/abs/0910.4183}{{\normalfont\ttfamily
  arXiv:0910.4183}}\relax
\mciteBstWouldAddEndPuncttrue
\mciteSetBstMidEndSepPunct{\mcitedefaultmidpunct}
{\mcitedefaultendpunct}{\mcitedefaultseppunct}\relax
\EndOfBibitem
\bibitem{Lansberg:2016deg}
J.-P. Lansberg and H.-S. Shao,
  \ifthenelse{\boolean{articletitles}}{\emph{{Towards an automated tool to
  evaluate the impact of the nuclear modification of the gluon density on
  quarkonium, D and B meson production in proton-nucleus collisions}},
  }{}\href{http://dx.doi.org/10.1140/epjc/s10052-016-4575-x}{Eur.\ Phys.\ J.\
  \textbf{C77} (2017) 1},
  \href{http://arxiv.org/abs/1610.05382}{{\normalfont\ttfamily
  arXiv:1610.05382}}\relax
\mciteBstWouldAddEndPuncttrue
\mciteSetBstMidEndSepPunct{\mcitedefaultmidpunct}
{\mcitedefaultendpunct}{\mcitedefaultseppunct}\relax
\EndOfBibitem
\bibitem{Shao:2015vga}
H.-S. Shao, \ifthenelse{\boolean{articletitles}}{\emph{{HELAC-Onia 2.0: an
  upgraded matrix-element and event generator for heavy quarkonium physics}},
  }{}\href{http://dx.doi.org/10.1016/j.cpc.2015.09.011}{Comput.\ Phys.\
  Commun.\  \textbf{198} (2016) 238},
  \href{http://arxiv.org/abs/1507.03435}{{\normalfont\ttfamily
  arXiv:1507.03435}}\relax
\mciteBstWouldAddEndPuncttrue
\mciteSetBstMidEndSepPunct{\mcitedefaultmidpunct}
{\mcitedefaultendpunct}{\mcitedefaultseppunct}\relax
\EndOfBibitem
\bibitem{Shao:2012iz}
H.-S. Shao, \ifthenelse{\boolean{articletitles}}{\emph{{HELAC-Onia: An
  automatic matrix element generator for heavy quarkonium physics}},
  }{}\href{http://dx.doi.org/10.1016/j.cpc.2013.05.023}{Comput.\ Phys.\
  Commun.\  \textbf{184} (2013) 2562},
  \href{http://arxiv.org/abs/1212.5293}{{\normalfont\ttfamily
  arXiv:1212.5293}}\relax
\mciteBstWouldAddEndPuncttrue
\mciteSetBstMidEndSepPunct{\mcitedefaultmidpunct}
{\mcitedefaultendpunct}{\mcitedefaultseppunct}\relax
\EndOfBibitem
\bibitem{nCTEQ15}
K.~Kovarik {\em et~al.}, \ifthenelse{\boolean{articletitles}}{\emph{{nCTEQ15 -
  Global analysis of nuclear parton distributions with uncertainties in the
  CTEQ framework}},
  }{}\href{http://dx.doi.org/10.1103/PhysRevD.93.085037}{Phys.\ Rev.\
  \textbf{D93} (2016) 085037},
  \href{http://arxiv.org/abs/1509.00792}{{\normalfont\ttfamily
  arXiv:1509.00792}}\relax
\mciteBstWouldAddEndPuncttrue
\mciteSetBstMidEndSepPunct{\mcitedefaultmidpunct}
{\mcitedefaultendpunct}{\mcitedefaultseppunct}\relax
\EndOfBibitem
\bibitem{Vogt}
R.~Vogt, \ifthenelse{\boolean{articletitles}}{\emph{Limits on intrinsic charm
  production from the SeaQuest experiment},
  }{}\href{http://dx.doi.org/10.1103/PhysRevC.103.035204}{Phys.\ Rev.\
  \textbf{C103} (2021) 035204}\relax
\mciteBstWouldAddEndPuncttrue
\mciteSetBstMidEndSepPunct{\mcitedefaultmidpunct}
{\mcitedefaultendpunct}{\mcitedefaultseppunct}\relax
\EndOfBibitem
\bibitem{NA51PLB438}
NA51 collaboration, M.~C. Abreu {\em et~al.},
  \ifthenelse{\boolean{articletitles}}{\emph{\jpsi, $\psi'$ and {Drell-Yan}
  production in pp and pd interactions at 450 {GeV/c}},
  }{}\href{http://dx.doi.org/https://doi.org/10.1016/S0370-2693(98)01014-4}{Phys.\
  Lett.\  \textbf{B438} (1998) 35}\relax
\mciteBstWouldAddEndPuncttrue
\mciteSetBstMidEndSepPunct{\mcitedefaultmidpunct}
{\mcitedefaultendpunct}{\mcitedefaultseppunct}\relax
\EndOfBibitem
\bibitem{NA50EPJC33}
NA50 collaboration, B.~Alessandro {\em et~al.},
  \ifthenelse{\boolean{articletitles}}{\emph{Charmonium production and nuclear
  absorption in {p-A} interactions at 450 {GeV}},
  }{}\href{http://dx.doi.org/10.1140/epjc/s2003-01539-y}{Eur.\ Phys.\ J.\
  \textbf{C33} (2004) 31}\relax
\mciteBstWouldAddEndPuncttrue
\mciteSetBstMidEndSepPunct{\mcitedefaultmidpunct}
{\mcitedefaultendpunct}{\mcitedefaultseppunct}\relax
\EndOfBibitem
\bibitem{NA50EPJC48}
NA50 collaboration, B.~Alessandro {\em et~al.},
  \ifthenelse{\boolean{articletitles}}{\emph{\jpsi and $\psi'$ production and
  their normal nuclear absorption in proton-nucleus collisions at 400 {GeV}},
  }{}\href{http://dx.doi.org/10.1140/epjc/s10052-006-0079-4}{Eur.\ Phys.\ J.\
  \textbf{C48} (2006) 329},
  \href{http://arxiv.org/abs/nucl-ex/0612012}{{\normalfont\ttfamily
  arXiv:nucl-ex/0612012}}\relax
\mciteBstWouldAddEndPuncttrue
\mciteSetBstMidEndSepPunct{\mcitedefaultmidpunct}
{\mcitedefaultendpunct}{\mcitedefaultseppunct}\relax
\EndOfBibitem
\bibitem{E771PLB374}
E771 collaboration, T.~Alexopoulos {\em et~al.},
  \ifthenelse{\boolean{articletitles}}{\emph{Production of \jpsi, $\psi'$ and
  {$\Upsilonres$} in 800 {GeV/c} proton-silicon interactions},
  }{}\href{http://dx.doi.org/10.1016/0370-2693(96)00256-0}{Phys.\ Lett.\
  \textbf{B374} (1996) 271}\relax
\mciteBstWouldAddEndPuncttrue
\mciteSetBstMidEndSepPunct{\mcitedefaultmidpunct}
{\mcitedefaultendpunct}{\mcitedefaultseppunct}\relax
\EndOfBibitem
\bibitem{E789PRD52}
E789 collaboration, M.~H. Schub {\em et~al.},
  \ifthenelse{\boolean{articletitles}}{\emph{Measurement of \jpsi and $\psi'$
  production for 800 {GeV/c} proton-gold collisions},
  }{}\href{http://dx.doi.org/10.1103/PhysRevD.52.1307}{Phys.\ Rev.\
  \textbf{D52} (1995) 1307}\relax
\mciteBstWouldAddEndPuncttrue
\mciteSetBstMidEndSepPunct{\mcitedefaultmidpunct}
{\mcitedefaultendpunct}{\mcitedefaultseppunct}\relax
\EndOfBibitem
\end{mcitethebibliography}
\end{document}